
%
%
%
%
%
%
%
\def\standardrisposta{s }\def\reducedrisposta{r }
\def\mplarisposta{mpla }\def\zerorisposta{z }
\def\doublerisposta{d }\def\cartarisposta{e }\def\amsrisposta{y }
\newcount\ingrandimento \newcount\sinnota \newcount\dimnota
\newcount\unoduecol \newdimen\collhsize \newdimen\tothsize
\newdimen\fullhsize \newcount\controllorisposta \sinnota=1
\newskip\infralinea  \global\controllorisposta=0
\immediate\write16 { ********  Welcome to PANDA macros (Plain TeX,
AP, 1991) ******** }
\immediate\write16 { You'll have to answer a few questions in
lowercase.}
\message{>  Do you want it in double-page (d), reduced (r)
or standard format (s) ? }\read-1 to\risposta
\message{>  Do you want it in USA A4 (u) or EUROPEAN A4
(e) paper size ? }\read-1 to\srisposta
\message{>  Do you have AMSFonts 2.0 (math) fonts (y/n) ? }
\read-1 to\arisposta
%
%
%
%
%
\ifx\risposta\standardrisposta \ingrandimento=1200
\message {>> This will come out UNREDUCED << }
\dimnota=2 \unoduecol=1 \global\controllorisposta=1 \fi
\ifx\risposta\reducedrisposta \ingrandimento=1095 \dimnota=1
\unoduecol=1  \global\controllorisposta=1
\message {>> This will come out REDUCED << } \fi
\ifx\risposta\doublerisposta \ingrandimento=1000 \dimnota=2
\unoduecol=2   \message {>> You must print this in
LANDSCAPE orientation << } \global\controllorisposta=1 \fi
\ifx\risposta\mplarisposta \ingrandimento=1000 \dimnota=1
\message {>> Mod. Phys. Lett. A format << }
\unoduecol=1 \global\controllorisposta=1 \fi
\ifx\risposta\zerorisposta \ingrandimento=1000 \dimnota=2
\message {>> Zero Magnification format << }
\unoduecol=1 \global\controllorisposta=1 \fi
\ifnum\controllorisposta=0  \ingrandimento=1200
\message {>>> ERROR IN INPUT, I ASSUME STANDARD
UNREDUCED FORMAT <<< }  \dimnota=2 \unoduecol=1 \fi
\magnification=\ingrandimento
%
%
%
%
\newdimen\eucolumnsize \newdimen\eudoublehsize \newdimen\eudoublevsize
\newdimen\uscolumnsize \newdimen\usdoublehsize \newdimen\usdoublevsize
\newdimen\eusinglehsize \newdimen\eusinglevsize \newdimen\ussinglehsize
\newskip\standardbaselineskip \newdimen\ussinglevsize
\newskip\reducedbaselineskip \newskip\doublebaselineskip
\eucolumnsize=12.0truecm    
\eudoublehsize=25.5truecm   
\eudoublevsize=6.5truein    
\uscolumnsize=4.4truein     
\usdoublehsize=9.4truein    
\usdoublevsize=6.8truein    
\eusinglehsize=6.5truein    
\eusinglevsize=24truecm     
\ussinglehsize=6.5truein    
\ussinglevsize=8.9truein    
\standardbaselineskip=16pt plus.2pt  
\reducedbaselineskip=14pt plus.2pt   
\doublebaselineskip=12pt plus.2pt    
%
%
\def\Portoffset{}
\def\Landoffset{}
\ifx\risposta\mplarisposta \def\Portoffset{\hoffset=1.8truecm} \fi
%
%
\def\Landspec{}
\tolerance=10000
\parskip=0pt plus2pt  \leftskip=0pt \rightskip=0pt
%
%
\ifx\risposta\standardrisposta \infralinea=\standardbaselineskip \fi
\ifx\risposta\reducedrisposta  \infralinea=\reducedbaselineskip \fi
\ifx\risposta\doublerisposta   \infralinea=\doublebaselineskip \fi
\ifx\risposta\mplarisposta     \infralinea=13pt \fi
\ifx\risposta\zerorisposta     \infralinea=12pt plus.2pt\fi
\ifnum\controllorisposta=0    \infralinea=\standardbaselineskip \fi
\ifx\risposta\doublerisposta   \Landoffset \else \Portoffset \fi
\ifx\risposta\doublerisposta \ifx\srisposta\cartarisposta
\tothsize=\eudoublehsize \collhsize=\eucolumnsize
\vsize=\eudoublevsize  \else  \tothsize=\usdoublehsize
\collhsize=\uscolumnsize \vsize=\usdoublevsize \fi \else
\ifx\srisposta\cartarisposta \tothsize=\eusinglehsize
\vsize=\eusinglevsize \else  \tothsize=\ussinglehsize
\vsize=\ussinglevsize \fi \collhsize=4.4truein \fi
\ifx\risposta\mplarisposta \tothsize=5.0truein
\vsize=7.8truein \collhsize=4.4truein \fi
%
%
%
%
\newcount\contaeuler \newcount\contacyrill \newcount\contaams
\font\ninerm=cmr9  \font\eightrm=cmr8  \font\sixrm=cmr6
\font\ninei=cmmi9  \font\eighti=cmmi8  \font\sixi=cmmi6
\font\ninesy=cmsy9  \font\eightsy=cmsy8  \font\sixsy=cmsy6
\font\ninebf=cmbx9  \font\eightbf=cmbx8  \font\sixbf=cmbx6
\font\ninett=cmtt9  \font\eighttt=cmtt8  \font\nineit=cmti9
\font\eightit=cmti8 \font\ninesl=cmsl9  \font\eightsl=cmsl8
\skewchar\ninei='177 \skewchar\eighti='177 \skewchar\sixi='177
\skewchar\ninesy='60 \skewchar\eightsy='60 \skewchar\sixsy='60
\hyphenchar\ninett=-1 \hyphenchar\eighttt=-1 \hyphenchar\tentt=-1
%
\font\tencmmib=cmmib10  \newfam\cmmibfam  \skewchar\tencmmib='177
\font\tencmbsy=cmbsy10  \newfam\cmbsyfam  \skewchar\tencmbsy='60
\def\scaps{\cmcsc}                 
\font\tencmcsc=cmcsc10  \newfam\cmcscfam
\ifnum\ingrandimento=1095

\font\capsone=cmcsc10 at 10.95pt 

\else

\font\capsone=cmcsc10 at 12pt 
\fi

\def\ttaarr{\bf}		
\def\ppaarr{\sl}		

%
%
%
\newfam\eufmfam \newfam\msamfam \newfam\msbmfam \newfam\eufbfam
\def\Loadeulerfonts{\global\contaeuler=1 \ifx\arisposta\amsrisposta
\font\teneufm=eufm10              
\font\eighteufm=eufm8 \font\nineeufm=eufm9 \font\sixeufm=eufm6
\font\seveneufm=eufm7  \font\fiveeufm=eufm5
\font\teneufb=eufb10              
\font\eighteufb=eufb8 \font\nineeufb=eufb9 \font\sixeufb=eufb6
\font\seveneufb=eufb7  \font\fiveeufb=eufb5
\font\teneurm=eurm10              
\font\eighteurm=eurm8 \font\nineeurm=eurm9
\font\teneurb=eurb10              
\font\eighteurb=eurb8 \font\nineeurb=eurb9
\font\teneusm=eusm10              
\font\eighteusm=eusm8 \font\nineeusm=eusm9
\font\teneusb=eusb10              
\font\eighteusb=eusb8 \font\nineeusb=eusb9
\else \def\eufm{\tt} \def\eufb{\tt} \def\eurm{\tt} \def\eurb{\tt}
\def\eusm{\tt} \def\eusb{\tt}    \fi}

\def\loadamsmath{\global\contaams=1 \ifx\arisposta\amsrisposta
\font\tenmsam=msam10 \font\ninemsam=msam9 \font\eightmsam=msam8
\font\sevenmsam=msam7 \font\sixmsam=msam6 \font\fivemsam=msam5
\font\tenmsbm=msbm10 \font\ninemsbm=msbm9 \font\eightmsbm=msbm8
\font\sevenmsbm=msbm7 \font\sixmsbm=msbm6 \font\fivemsbm=msbm5
\else \def\msbm{\bf} \fi \def\Bbb{\msbm} \def\symbl{\msam} \tenpoint}
\def\loadcyrill{\global\contacyrill=1 \ifx\arisposta\amsrisposta
\font\tenwncyr=wncyr10 \font\ninewncyr=wncyr9 \font\eightwncyr=wncyr8
\font\tenwncyb=wncyr10 \font\ninewncyb=wncyr9 \font\eightwncyb=wncyr8
\font\tenwncyi=wncyr10 \font\ninewncyi=wncyr9 \font\eightwncyi=wncyr8
\else \def\cyrill{\sl} \def\cyrilb{\sl} \def\cyrili{\sl} \fi\tenpoint}
\ifx\arisposta\amsrisposta
\font\sevenex=cmex7               
\font\eightex=cmex8  \font\nineex=cmex9
\font\ninecmmib=cmmib9   \font\eightcmmib=cmmib8
\font\sevencmmib=cmmib7 \font\sixcmmib=cmmib6
\font\fivecmmib=cmmib5   \skewchar\ninecmmib='177
\skewchar\eightcmmib='177  \skewchar\sevencmmib='177
\skewchar\sixcmmib='177   \skewchar\fivecmmib='177
\font\ninecmbsy=cmbsy9    \font\eightcmbsy=cmbsy8
\font\sevencmbsy=cmbsy7  \font\sixcmbsy=cmbsy6
\font\fivecmbsy=cmbsy5   \skewchar\ninecmbsy='60
\skewchar\eightcmbsy='60  \skewchar\sevencmbsy='60
\skewchar\sixcmbsy='60    \skewchar\fivecmbsy='60
\font\ninecmcsc=cmcsc9    \font\eightcmcsc=cmcsc8     \else
\def\cmmib{\fam\cmmibfam\tencmmib}\textfont\cmmibfam=\tencmmib
\scriptfont\cmmibfam=\tencmmib \scriptscriptfont\cmmibfam=\tencmmib
\def\cmbsy{\fam\cmbsyfam\tencmbsy} \textfont\cmbsyfam=\tencmbsy
\scriptfont\cmbsyfam=\tencmbsy \scriptscriptfont\cmbsyfam=\tencmbsy
\scriptfont\cmcscfam=\tencmcsc \scriptscriptfont\cmcscfam=\tencmcsc
\def\cmcsc{\fam\cmcscfam\tencmcsc} \textfont\cmcscfam=\tencmcsc \fi
\catcode`@=11
\newskip\ttglue
\gdef\tenpoint{\def\rm{\fam0\tenrm}
  \textfont0=\tenrm \scriptfont0=\sevenrm \scriptscriptfont0=\fiverm
  \textfont1=\teni \scriptfont1=\seveni \scriptscriptfont1=\fivei
  \textfont2=\tensy \scriptfont2=\sevensy \scriptscriptfont2=\fivesy
  \textfont3=\tenex \scriptfont3=\tenex \scriptscriptfont3=\tenex
  \def\mcal{\fam2 \tensy}  \def\mmit{\fam1 \teni}
  \textfont\itfam=\tenit \def\it{\fam\itfam\tenit}
  \textfont\slfam=\tensl \def\sl{\fam\slfam\tensl}
  \textfont\ttfam=\tentt \scriptfont\ttfam=\eighttt
  \scriptscriptfont\ttfam=\eighttt  \def\tt{\fam\ttfam\tentt}
  \textfont\bffam=\tenbf \scriptfont\bffam=\sevenbf
  \scriptscriptfont\bffam=\fivebf \def\bf{\fam\bffam\tenbf}
     \ifx\arisposta\amsrisposta    \ifnum\contaeuler=1
  \textfont\eufmfam=\teneufm \scriptfont\eufmfam=\seveneufm
  \scriptscriptfont\eufmfam=\fiveeufm \def\eufm{\fam\eufmfam\teneufm}
  \textfont\eufbfam=\teneufb \scriptfont\eufbfam=\seveneufb
  \scriptscriptfont\eufbfam=\fiveeufb \def\eufb{\fam\eufbfam\teneufb}
  \def\eurm{\teneurm} \def\eurb{\teneurb} \def\eusm{\teneusm}
  \def\eusb{\teneusb}    \fi    \ifnum\contaams=1
  \textfont\msamfam=\tenmsam \scriptfont\msamfam=\sevenmsam
  \scriptscriptfont\msamfam=\fivemsam \def\msam{\fam\msamfam\tenmsam}
  \textfont\msbmfam=\tenmsbm \scriptfont\msbmfam=\sevenmsbm
  \scriptscriptfont\msbmfam=\fivemsbm \def\msbm{\fam\msbmfam\tenmsbm}
     \fi      \ifnum\contacyrill=1     \def\cyrill{\tenwncyr}
  \def\cyrilb{\tenwncyb}  \def\cyrili{\tenwncyi}         \fi
  \textfont3=\tenex \scriptfont3=\sevenex \scriptscriptfont3=\sevenex
  \def\cmmib{\fam\cmmibfam\tencmmib} \scriptfont\cmmibfam=\sevencmmib
  \textfont\cmmibfam=\tencmmib  \scriptscriptfont\cmmibfam=\fivecmmib
  \def\cmbsy{\fam\cmbsyfam\tencmbsy} \scriptfont\cmbsyfam=\sevencmbsy
  \textfont\cmbsyfam=\tencmbsy  \scriptscriptfont\cmbsyfam=\fivecmbsy
  \def\cmcsc{\fam\cmcscfam\tencmcsc} \scriptfont\cmcscfam=\eightcmcsc
  \textfont\cmcscfam=\tencmcsc \scriptscriptfont\cmcscfam=\eightcmcsc
     \fi            \tt \ttglue=.5em plus.25em minus.15em
  \normalbaselineskip=12pt
  \setbox\strutbox=\hbox{\vrule height8.5pt depth3.5pt width0pt}
  \let\sc=\eightrm \let\big=\tenbig   \normalbaselines
  \baselineskip=\infralinea  \rm}
\gdef\ninepoint{\def\rm{\fam0\ninerm}
  \textfont0=\ninerm \scriptfont0=\sixrm \scriptscriptfont0=\fiverm
  \textfont1=\ninei \scriptfont1=\sixi \scriptscriptfont1=\fivei
  \textfont2=\ninesy \scriptfont2=\sixsy \scriptscriptfont2=\fivesy
  \textfont3=\tenex \scriptfont3=\tenex \scriptscriptfont3=\tenex
  \def\mcal{\fam2 \ninesy}  \def\mmit{\fam1 \ninei}
  \textfont\itfam=\nineit \def\it{\fam\itfam\nineit}
  \textfont\slfam=\ninesl \def\sl{\fam\slfam\ninesl}
  \textfont\ttfam=\ninett \scriptfont\ttfam=\eighttt
  \scriptscriptfont\ttfam=\eighttt \def\tt{\fam\ttfam\ninett}
  \textfont\bffam=\ninebf \scriptfont\bffam=\sixbf
  \scriptscriptfont\bffam=\fivebf \def\bf{\fam\bffam\ninebf}
     \ifx\arisposta\amsrisposta  \ifnum\contaeuler=1
  \textfont\eufmfam=\nineeufm \scriptfont\eufmfam=\sixeufm
  \scriptscriptfont\eufmfam=\fiveeufm \def\eufm{\fam\eufmfam\nineeufm}
  \textfont\eufbfam=\nineeufb \scriptfont\eufbfam=\sixeufb
  \scriptscriptfont\eufbfam=\fiveeufb \def\eufb{\fam\eufbfam\nineeufb}
  \def\eurm{\nineeurm} \def\eurb{\nineeurb} \def\eusm{\nineeusm}
  \def\eusb{\nineeusb}     \fi   \ifnum\contaams=1
  \textfont\msamfam=\ninemsam \scriptfont\msamfam=\sixmsam
  \scriptscriptfont\msamfam=\fivemsam \def\msam{\fam\msamfam\ninemsam}
  \textfont\msbmfam=\ninemsbm \scriptfont\msbmfam=\sixmsbm
  \scriptscriptfont\msbmfam=\fivemsbm \def\msbm{\fam\msbmfam\ninemsbm}
     \fi       \ifnum\contacyrill=1     \def\cyrill{\ninewncyr}
  \def\cyrilb{\ninewncyb}  \def\cyrili{\ninewncyi}         \fi
  \textfont3=\nineex \scriptfont3=\sevenex \scriptscriptfont3=\sevenex
  \def\cmmib{\fam\cmmibfam\ninecmmib}  \textfont\cmmibfam=\ninecmmib
  \scriptfont\cmmibfam=\sixcmmib \scriptscriptfont\cmmibfam=\fivecmmib
  \def\cmbsy{\fam\cmbsyfam\ninecmbsy}  \textfont\cmbsyfam=\ninecmbsy
  \scriptfont\cmbsyfam=\sixcmbsy \scriptscriptfont\cmbsyfam=\fivecmbsy
  \def\cmcsc{\fam\cmcscfam\ninecmcsc} \scriptfont\cmcscfam=\eightcmcsc
  \textfont\cmcscfam=\ninecmcsc \scriptscriptfont\cmcscfam=\eightcmcsc
     \fi            \tt \ttglue=.5em plus.25em minus.15em
  \normalbaselineskip=11pt
  \setbox\strutbox=\hbox{\vrule height8pt depth3pt width0pt}
  \let\sc=\sevenrm \let\big=\ninebig \normalbaselines\rm}
\gdef\eightpoint{\def\rm{\fam0\eightrm}
  \textfont0=\eightrm \scriptfont0=\sixrm \scriptscriptfont0=\fiverm
  \textfont1=\eighti \scriptfont1=\sixi \scriptscriptfont1=\fivei
  \textfont2=\eightsy \scriptfont2=\sixsy \scriptscriptfont2=\fivesy
  \textfont3=\tenex \scriptfont3=\tenex \scriptscriptfont3=\tenex
  \def\mcal{\fam2 \eightsy}  \def\mmit{\fam1 \eighti}
  \textfont\itfam=\eightit \def\it{\fam\itfam\eightit}
  \textfont\slfam=\eightsl \def\sl{\fam\slfam\eightsl}
  \textfont\ttfam=\eighttt \scriptfont\ttfam=\eighttt
  \scriptscriptfont\ttfam=\eighttt \def\tt{\fam\ttfam\eighttt}
  \textfont\bffam=\eightbf \scriptfont\bffam=\sixbf
  \scriptscriptfont\bffam=\fivebf \def\bf{\fam\bffam\eightbf}
     \ifx\arisposta\amsrisposta   \ifnum\contaeuler=1
  \textfont\eufmfam=\eighteufm \scriptfont\eufmfam=\sixeufm
  \scriptscriptfont\eufmfam=\fiveeufm \def\eufm{\fam\eufmfam\eighteufm}
  \textfont\eufbfam=\eighteufb \scriptfont\eufbfam=\sixeufb
  \scriptscriptfont\eufbfam=\fiveeufb \def\eufb{\fam\eufbfam\eighteufb}
  \def\eurm{\eighteurm} \def\eurb{\eighteurb} \def\eusm{\eighteusm}
  \def\eusb{\eighteusb}       \fi    \ifnum\contaams=1
  \textfont\msamfam=\eightmsam \scriptfont\msamfam=\sixmsam
  \scriptscriptfont\msamfam=\fivemsam \def\msam{\fam\msamfam\eightmsam}
  \textfont\msbmfam=\eightmsbm \scriptfont\msbmfam=\sixmsbm
  \scriptscriptfont\msbmfam=\fivemsbm \def\msbm{\fam\msbmfam\eightmsbm}
     \fi       \ifnum\contacyrill=1     \def\cyrill{\eightwncyr}
  \def\cyrilb{\eightwncyb}  \def\cyrili{\eightwncyi}         \fi
  \textfont3=\eightex \scriptfont3=\sevenex \scriptscriptfont3=\sevenex
  \def\cmmib{\fam\cmmibfam\eightcmmib}  \textfont\cmmibfam=\eightcmmib
  \scriptfont\cmmibfam=\sixcmmib \scriptscriptfont\cmmibfam=\fivecmmib
  \def\cmbsy{\fam\cmbsyfam\eightcmbsy}  \textfont\cmbsyfam=\eightcmbsy
  \scriptfont\cmbsyfam=\sixcmbsy \scriptscriptfont\cmbsyfam=\fivecmbsy
  \def\cmcsc{\fam\cmcscfam\eightcmcsc} \scriptfont\cmcscfam=\eightcmcsc
  \textfont\cmcscfam=\eightcmcsc \scriptscriptfont\cmcscfam=\eightcmcsc
     \fi             \tt \ttglue=.5em plus.25em minus.15em
  \normalbaselineskip=9pt
  \setbox\strutbox=\hbox{\vrule height7pt depth2pt width0pt}
  \let\sc=\sixrm \let\big=\eightbig \normalbaselines\rm }
\gdef\tenbig#1{{\hbox{$\left#1\vbox to8.5pt{}\right.\n@space$}}}
\gdef\ninebig#1{{\hbox{$\textfont0=\tenrm\textfont2=\tensy
   \left#1\vbox to7.25pt{}\right.\n@space$}}}
\gdef\eightbig#1{{\hbox{$\textfont0=\ninerm\textfont2=\ninesy
   \left#1\vbox to6.5pt{}\right.\n@space$}}}
\def\alternativefont#1#2{\ifx\arisposta\amsrisposta \relax \else
\xdef#1{#2} \fi}
\global\contaeuler=0 \global\contacyrill=0 \global\contaams=0
%
%
%
%
\newbox\fotlinebb \newbox\hedlinebb \newbox\leftcolumn
\gdef\makeheadline{\vbox to 0pt{\vskip-22.5pt
     \fullline{\vbox to8.5pt{}\the\headline}\vss}\nointerlineskip}
\gdef\makehedlinebb{\vbox to 0pt{\vskip-22.5pt
     \fullline{\vbox to8.5pt{}\copy\hedlinebb\hfil
     \line{\hfill\the\headline\hfill}}\vss} \nointerlineskip}
\gdef\makefootline{\baselineskip=24pt \fullline{\the\footline}}
\gdef\makefotlinebb{\baselineskip=24pt
    \fullline{\copy\fotlinebb\hfil\line{\hfill\the\footline\hfill}}}
\gdef\doubleformat{\shipout\vbox{\Landspec\makehedlinebb
     \fullline{\box\leftcolumn\hfil\columnbox}\makefotlinebb}
     \advancepageno}
\gdef\columnbox{\leftline{\pagebody}}
\gdef\line#1{\hbox to\hsize{\hskip\leftskip#1\hskip\rightskip}}
\gdef\fullline#1{\hbox to\fullhsize{\hskip\leftskip{#1}%
\hskip\rightskip}}
\gdef\footnote#1{\let\@sf=\empty
         \ifhmode\edef\#sf{\spacefactor=\the\spacefactor}\/\fi
         #1\@sf\vfootnote{#1}}
\gdef\vfootnote#1{\insert\footins\bgroup
         \ifnum\dimnota=1  \eightpoint\fi
         \ifnum\dimnota=2  \ninepoint\fi
         \ifnum\dimnota=0  \tenpoint\fi
         \interlinepenalty=\interfootnotelinepenalty
         \splittopskip=\ht\strutbox
         \splitmaxdepth=\dp\strutbox \floatingpenalty=20000
         \leftskip=\oldssposta \rightskip=\olddsposta
         \spaceskip=0pt \xspaceskip=0pt
         \ifnum\sinnota=0   \textindent{#1}\fi
         \ifnum\sinnota=1   \item{#1}\fi
         \footstrut\futurelet\next\fo@t}
\gdef\fo@t{\ifcat\bgroup\noexpand\next \let\next\f@@t
             \else\let\next\f@t\fi \next}
\gdef\f@@t{\bgroup\aftergroup\@foot\let\next}
\gdef\f@t#1{#1\@foot} \gdef\@foot{\strut\egroup}
\gdef\footstrut{\vbox to\splittopskip{}}
\skip\footins=\bigskipamount
\count\footins=1000  \dimen\footins=8in
\catcode`@=12
\tenpoint
\ifnum\unoduecol=1 \hsize=\tothsize   \fullhsize=\tothsize \fi
\ifnum\unoduecol=2 \hsize=\collhsize  \fullhsize=\tothsize \fi
\global\let\lrcol=L      \ifnum\unoduecol=1
\output{\plainoutput{\ifnum\tipbnota=2 \clearnmbnota\fi}} \fi
\ifnum\unoduecol=2 \output{\if L\lrcol
     \global\setbox\leftcolumn=\columnbox
     \global\setbox\fotlinebb=\line{\hfill\the\footline\hfill}
     \global\setbox\hedlinebb=\line{\hfill\the\headline\hfill}
     \advancepageno  \global\let\lrcol=R
     \else  \doubleformat \global\let\lrcol=L \fi
     \ifnum\outputpenalty>-20000 \else\dosupereject\fi
     \ifnum\tipbnota=2\clearnmbnota\fi }\fi
\def\ifdoublepage{\ifnum\unoduecol=2 }
\gdef\yespagenumbers{\footline={\hss\tenrm\folio\hss}}
\gdef\ciao{ \ifnum\fdefcontre=1 \endfdef\fi
     \par\vfill\supereject \ifnum\unoduecol=2
     \if R\lrcol  \headline={}\nopagenumbers\null\vfill\eject
     \fi\fi \end}

\newskip\olddsposta \newskip\oldssposta
\global\oldssposta=\leftskip \global\olddsposta=\rightskip

\def\filldots{\leaders\hbox to 1em{\hss.\hss}\hfill}
\def\inquadrb#1 {\vbox {\hrule  \hbox{\vrule \vbox {\vskip .2cm
    \hbox {\ #1\ } \vskip .2cm } \vrule  }  \hrule} }
 \def\newline{\hfil\break}
\def\jump{\vskip\baselineskip} \newskip\iinnffrr
\def\sjump{\iinnffrr=\baselineskip
          \divide\iinnffrr by 2 \vskip\iinnffrr}
\def\bjump{\vskip\baselineskip \vskip\baselineskip}
\newcount\nmbnota  \def\clearnmbnota{\global\nmbnota=0}
\newcount\tipbnota \def\letterfootnote{\global\tipbnota=1}

\def\note#1{\global\advance\nmbnota by 1 \ifnum\tipbnota=1
    \footnote{$^{\rm\nttlett}$}{#1} \else {\ifnum\tipbnota=2
    \footnote{$^{\nttsymb}$}{#1}
    \else\footnote{$^{\the\nmbnota}$}{#1}\fi}\fi}
\def\nttlett{\ifcase\nmbnota \or a\or b\or c\or d\or e\or f\or
g\or h\or i\or j\or k\or l\or m\or n\or o\or p\or q\or r\or
s\or t\or u\or v\or w\or y\or x\or z\fi}
\def\nttsymb{\ifcase\nmbnota \or\dag\or\sharp\or\ddag\or\star\or
\natural\or\flat\or\clubsuit\or\diamondsuit\or\heartsuit
\or\spadesuit\fi}   \clearnmbnota
\def\numberfootnote{\global\tipbnota=0} \numberfootnote
\def\setnote#1{\expandafter\xdef\csname#1\endcsname{
\ifnum\tipbnota=1 {\rm\nttlett} \else {\ifnum\tipbnota=2
{\nttsymb} \else \the\nmbnota\fi}\fi} }
\newcount\nbmfig  \def\clearnbmfig{\global\nbmfig=0}
\gdef\figure{\global\advance\nbmfig by 1
      {\rm fig. \the\nbmfig}}   \clearnbmfig
\def\setfig#1{\expandafter\xdef\csname#1\endcsname{fig. \the\nbmfig}}
 \def\endformula{\eqno\numero $$}
 \def\efr{\endformula}
\newcount\frmcount \def\clearfrmcount{\global\frmcount=0}
\def\numero{\global\advance\frmcount by 1   \ifnum\indappcount=0
  {\ifnum\cpcount <1 {\hbox{\rm (\the\frmcount )}}  \else
  {\hbox{\rm (\the\cpcount .\the\frmcount )}} \fi}  \else
  {\hbox{\rm (\applett .\the\frmcount )}} \fi}
\def\nameformula#1{\global\advance\frmcount by 1%
\ifnum\draftnum=0  {\ifnum\indappcount=0%
{\ifnum\cpcount<1\xdef\spzzttrra{(\the\frmcount )}%
\else\xdef\spzzttrra{(\the\cpcount .\the\frmcount )}\fi}%
\else\xdef\spzzttrra{(\applett .\the\frmcount )}\fi}%
\else\xdef\spzzttrra{(#1)}\fi%
\expandafter\xdef\csname#1\endcsname{\spzzttrra}
\eqno \hbox{\rm\spzzttrra} $$}
\def\nfr{\nameformula}    
\def\nameali#1{\global\advance\frmcount by 1%
\ifnum\draftnum=0  {\ifnum\indappcount=0%
{\ifnum\cpcount<1\xdef\spzzttrra{(\the\frmcount )}%
\else\xdef\spzzttrra{(\the\cpcount .\the\frmcount )}\fi}%
\else\xdef\spzzttrra{(\applett .\the\frmcount )}\fi}%
\else\xdef\spzzttrra{(#1)}\fi%
\expandafter\xdef\csname#1\endcsname{\spzzttrra}
  \hbox{\rm\spzzttrra} }      \clearfrmcount
\newcount\cpcount \def\clearcpcount{\global\cpcount=0}
\newcount\subcpcount \def\clearsubcpcount{\global\subcpcount=0}
\newcount\appcount \def\clearappcount{\global\appcount=0}
\newcount\indappcount \def\clearindappcount{\indappcount=0}
\newcount\sottoparcount 

\def\applett{\ifcase\appcount  \or {A}\or {B}\or {C}\or
{D}\or {E}\or {F}\or {G}\or {H}\or {I}\or {J}\or {K}\or {L}\or
{M}\or {N}\or {O}\or {P}\or {Q}\or {R}\or {S}\or {T}\or {U}\or
{V}\or {W}\or {X}\or {Y}\or {Z}\fi    \ifnum\appcount<0
\immediate\write16 {Panda ERROR - Appendix: counter "appcount"
out of range}\fi  \ifnum\appcount>26  \immediate\write16 {Panda
ERROR - Appendix: counter "appcount" out of range}\fi}
\clearappcount  \clearindappcount \newcount\connttrre
\def\clearconnttrre{\global\connttrre=0} \newcount\countref
\def\clearcountref{\global\countref=0} \clearcountref
\def\chapter#1{\global\advance\cpcount by 1 \clearfrmcount
                 \goodbreak\null\vbox{\jump\nobreak
                 \clearsubcpcount\clearindappcount
                 \itemitem{\ttaarr\the\cpcount .\qquad}{\ttaarr #1}
                 \par\nobreak\jump\sjump}\nobreak}
\def\section#1{\global\advance\subcpcount by 1 \goodbreak\null
               \vbox{\sjump\nobreak\ifnum\indappcount=0
                 {\ifnum\cpcount=0 {\itemitem{\ppaarr
               .\the\subcpcount\quad\enskip\ }{\ppaarr #1}\par} \else
                 {\itemitem{\ppaarr\the\cpcount .\the\subcpcount\quad
                  \enskip\ }{\ppaarr #1} \par}  \fi}
                \else{\itemitem{\ppaarr\applett .\the\subcpcount\quad
                 \enskip\ }{\ppaarr #1}\par}\fi\nobreak\jump}\nobreak}
\clearsubcpcount
\def\appendix#1{\global\advance\appcount by 1 \clearfrmcount
                  \goodbreak\null\vbox{\jump\nobreak
                  \global\advance\indappcount by 1 \clearsubcpcount
          \itemitem{ }{\hskip-40pt\ttaarr Appendix\ \applett :\ #1}
             \nobreak\jump\sjump}\nobreak}
\clearappcount \clearindappcount
\def\references{\goodbreak\null\vbox{\jump\nobreak
   \itemitem{}{\ttaarr References} \nobreak\jump\sjump}\nobreak}

\clearcpcount\clearcountref

\def\setchap#1{\ifnum\indappcount=0{\ifnum\subcpcount=0%
\xdef\spzzttrra{\the\cpcount}%
\else\xdef\spzzttrra{\the\cpcount .\the\subcpcount}\fi}
\else{\ifnum\subcpcount=0 \xdef\spzzttrra{\applett}%
\else\xdef\spzzttrra{\applett .\the\subcpcount}\fi}\fi
\expandafter\xdef\csname#1\endcsname{\spzzttrra}}
\newcount\draftnum \newcount\ppora   \newcount\ppminuti
\global\ppora=\time   \global\ppminuti=\time
\global\divide\ppora by 60  \draftnum=\ppora
\multiply\draftnum by 60    \global\advance\ppminuti by -\draftnum
\def\droggi{\number\day /\number\month /\number\year\ \the\ppora
:\the\ppminuti}     \global\draftnum=0
\def\draftcomment#1{\ifnum\draftnum=0 \relax \else
{\ {\bf ***}\ #1\ {\bf ***}\ }\fi} 
%
%
\catcode`@=11
\gdef\Ref#1{\expandafter\ifx\csname @rrxx@#1\endcsname\relax%
{\global\advance\countref by 1    \ifnum\countref>200
\immediate\write16 {Panda ERROR - Ref: maximum number of references
exceeded}  \expandafter\xdef\csname @rrxx@#1\endcsname{0}\else
\expandafter\xdef\csname @rrxx@#1\endcsname{\the\countref}\fi}\fi
\ifnum\draftnum=0 \csname @rrxx@#1\endcsname \else#1\fi}
\gdef\beginref{\ifnum\draftnum=0  \gdef\Rref{\fairef}
\gdef\endref{\scriviref} \else\relax\fi
\ifx\risposta\mplarisposta \ninepoint \fi
\parskip 2pt plus.2pt \baselineskip=12pt}
\def\Reflab#1{[#1]} \gdef\Rref#1#2{\item{\Reflab{#1}}{#2}}
\gdef\endref{\relax}  \newcount\conttemp
\gdef\fairef#1#2{\expandafter\ifx\csname @rrxx@#1\endcsname\relax
{\global\conttemp=0 \immediate\write16 {Panda ERROR - Ref: reference
[#1] undefined}} \else
{\global\conttemp=\csname @rrxx@#1\endcsname } \fi
\global\advance\conttemp by 50  \global\setbox\conttemp=\hbox{#2} }
\gdef\scriviref{\clearconnttrre\conttemp=50
\loop\ifnum\connttrre<\countref \advance\conttemp by 1
\advance\connttrre by 1
\item{\Reflab{\the\connttrre}}{\unhcopy\conttemp} \repeat}
\clearcountref \clearconnttrre
\catcode`@=12
\ifx\risposta\mplarisposta \def\Reflab#1{#1.} \letterfootnote \fi

\def\slashchar#1{\setbox0=\hbox{$#1$} \dimen0=\wd0
     \setbox1=\hbox{/} \dimen1=\wd1 \ifdim\dimen0>\dimen1
      \rlap{\hbox to \dimen0{\hfil/\hfil}} #1 \else
      \rlap{\hbox to \dimen1{\hfil$#1$\hfil}} / \fi}
\ifx\oldchi\undefined \let\oldchi=\chi
  \def\cchi{{\raise 1pt\hbox{$\oldchi$}}} \let\chi=\cchi \fi

\def\frac#1#2{{\textstyle{#1 \over #2}}}

\def\half{\ifinner {\scriptstyle {1 \over 2}}\else {1 \over 2} \fi}

\def\simge{\rlap{\raise 2pt \hbox{$>$}}{\lower 2pt \hbox{$\sim$}}}
\def\simle{\rlap{\raise 2pt \hbox{$<$}}{\lower 2pt \hbox{$\sim$}}}

\def\vbig#1#2{{\vbigd@men=#2\divide\vbigd@men by 2%
\hbox{$\left#1\vbox to \vbigd@men{}\right.\n@space$}}}

%
%
\newcount\fdefcontre \newcount\fdefcount \newcount\indcount
\newread\filefdef  \newread\fileftmp  \newwrite\filefdef
\newwrite\fileftmp     \def\strip#1*.A {#1}
\def\futuredef#1{\beginfdef
\expandafter\ifx\csname#1\endcsname\relax%
{\immediate\write\fileftmp {#1*.A}
\immediate\write16 {Panda Warning - fdef: macro "#1" on page
\the\pageno \space undefined}
\ifnum\draftnum=0 \expandafter\xdef\csname#1\endcsname{(?)}
\else \expandafter\xdef\csname#1\endcsname{(#1)} \fi
\global\advance\fdefcount by 1}\fi   \csname#1\endcsname}

\def\beginfdef{\ifnum\fdefcontre=0
\immediate\openin\filefdef \jobname.fdef
\immediate\openout\fileftmp \jobname.ftmp
\global\fdefcontre=1  \ifeof\filefdef \immediate\write16 {Panda
WARNING - fdef: file \jobname.fdef not found, run TeX again}
\else \immediate\read\filefdef to\spzzttrra
\global\advance\fdefcount by \spzzttrra
\indcount=0      \loop\ifnum\indcount<\fdefcount
\advance\indcount by 1   \immediate\read\filefdef to\spezttrra
\immediate\read\filefdef to\sppzttrra
\edef\spzzttrra{\expandafter\strip\spezttrra}
\immediate\write\fileftmp {\spzzttrra *.A}
\expandafter\xdef\csname\spzzttrra\endcsname{\sppzttrra}
\repeat \fi \immediate\closein\filefdef \fi}
\def\endfdef{\immediate\closeout\fileftmp   \ifnum\fdefcount>0
\immediate\openin\fileftmp \jobname.ftmp
\immediate\openout\filefdef \jobname.fdef
\immediate\write\filefdef {\the\fdefcount}   \indcount=0
\loop\ifnum\indcount<\fdefcount    \advance\indcount by 1
\immediate\read\fileftmp to\spezttrra
\edef\spzzttrra{\expandafter\strip\spezttrra}
\immediate\write\filefdef{\spzzttrra *.A}
\edef\spezttrra{\string{\csname\spzzttrra\endcsname\string}}
\iwritel\filefdef{\spezttrra}
\repeat  \immediate\closein\fileftmp \immediate\closeout\filefdef
\immediate\write16 {Panda Warning - fdef: Label(s) may have changed,
re-run TeX to get them right}\fi}
\def\iwritel#1#2{\newlinechar=-1
{\newlinechar=`\ \immediate\write#1{#2}}\newlinechar=-1}
\global\fdefcontre=0 \global\fdefcount=0 \global\indcount=0
%
%
\null
%
%
%
%

%
\loadamsmath
\def\t{\theta}
%
\pageno=0\baselineskip=14pt
\nopagenumbers{
\line{\hfill CERN-TH.6949/93}
\line{\hfill\tt hep-th/9308147}
\line{\hfill SWAT/92-93/11}
\line{\hfill August 1993}
\ifdoublepage \bjump\bjump\bjump\bjump\else\vfill\fi
\centerline{\capsone FROM $A_{m-1}$ TRIGONOMETRIC S-MATRICES TO}
\sjump\sjump
\centerline{\capsone THE THERMODYNAMIC BETHE ANSATZ}
\bjump\bjump
\centerline{\scaps Timothy J. Hollowood\footnote{$^*$}{On leave from:
Department of Mathematics, University College of Swansea, SA2 8PP, U.K.} }
\sjump
\sjump
\centerline{\sl CERN-TH, 1211 Geneva 23, Switzerland.}
\centerline{\tt hollow@surya11.cern.ch}
\bjump\bjump\bjump
\ifdoublepage
\vfill
\noindent
\line{CERN-TH.6949/93\hfill}
\line{SWAT/92-93/11}
\line{August 1993\hfill}
\eject\null\vfill\fi
\centerline{\capsone ABSTRACT}\sjump
The thermodynamic Bethe Ansatz equations that have been proposed to
describe massive
integrable deformations of the coset conformal field theories
$g_k\times g_l/g_{k+l}$ are shown to result directly by applying the
usual thermodynamic Bethe Ansatz arguments to the trigonometric
$S$-matrices for the algebras $g=a_{m-1}$.
\sjump\vfill
\ifdoublepage \else
\noindent
\line{CERN-TH.6949/93\hfill}
\line{August 1993\hfill}\fi
\eject}
\yespagenumbers\pageno=1
%
%

\chapter{Introduction}

Thermodynamic Bethe Ansatz (TBA) systems of equations have been conjectured to
describe the renormalization group trajectories issuing in various
directions from certain coset conformal field theories.
Although perfectly reasonable, these conjectures, in most cases,
have not been shown to arise from a consistent factorizable $S$-matrix
theory; the exceptions being when the
particles carry no internal quantum numbers in which case
the $S$-matrix is purely elastic [\Ref{TBA}], and some examples
involving the algebra $a_1$ [\Ref{ND},\Ref{FI}]. In particular, TBA systems
have been written down [\Ref{RAV}] which should describe the
massive deformation of
the coset $g_k\times g_l/g_{k+l}$, for a simply-laced algebra $g$ (the
subscripts denoting the level) by the relevant operator known conventionally
as $\phi_{1,1,{\rm adj}}$
having conformal dimension $\Delta=(k+l)/(\tilde h+k+l)$,
where $\tilde h$ is the dual Coxeter number of $g$.

For the $a_{m-1}$ algebras strong
support for the proposed TBA systems comes in an entirely different
direction from the {\it restricted solid-on-solid\/} (RSOS) integrable lattice
models [\Ref{RSOS}], that have been solved using Bethe Ansatz techniques
[\Ref{BRI}], and which in the scaling limit are thought to be identical to
the quantum field theories describing the deformations of the cosets
discussed above [\Ref{BRI}]. In particular,
[\Ref{BRII}] concluded that the TBA systems resulting from the lattice
models should be associated to integrable quantum field theories with
factorizable $S$-matrices whose elements are proportional to RSOS Boltzmann
weights. The proposed $S$-matrices are the {\it trigonometric\/} $S$-matrices
associated to $a_{m-1}$. The $S$-matrix elements for the elementary states
were written down in [\Ref{BRII},\Ref{VF}] (see also [\Ref{ABL}]),
however, the bootstrap
was solved in [\Ref{THI}] and the full $S$-matrix written down explicitly in
[\Ref{THII}].

With the expression for the complete $S$-matrix it is now
possible to complete the picture and
verify that the TBA systems from the lattice models, and hence those
conjectured to describe the integrable deformations of the cosets, do
actually come directly from a consistent factorizable $S$-matrix.
Unfortunately, as with most other calculations involving the Bethe
Ansatz, it is not possible to be completely rigorous and some
reasonable conjectures must be made; for instance, the use of the
string hypothesis.

\chapter{The trigonometric S-matrices}

In this section we write down
the trigonometric S-matrices associated to the Lie algebra $a_{m-1}$.
A full discussion of these $S$-matrices and a proof of the fact that they
satisfy all the necessary requirements may be found in [\Ref{THI},\Ref{THII}].

The particles transform in representations $V_a\otimes
V_a$, $a=1,2,\ldots,m-1$, of the tensor
product $a_{m-1}\times a_{m-1}$ ($V_a$ are the fundamental representations
of $a_{m-1}$ whose highest weights are the fundamental weights $\omega_a$).
The masses of the multiplets are
$$
m_a=m_0\sin\left({\pi a\over m}\right),\qquad a=1,2,\ldots,m-1,
\efr
where $m_0$ is some overall mass scale. The $S$-matrix depends upon two
additional parameters $k$ and $l$ which in this paper will be taken to be
natural numbers but more generally can be thought of as coupling constants.

The two-body $S$-matrix element between the $a^{\rm th}$ and
$b^{\rm th}$ multiplet has the following factored structure:
$$
S^{ab}_{(k,l)}(\theta)=X^{ab}(\theta){\widetilde S}^{ab}_{(k)}(\theta)
\otimes{\widetilde S}^{ab}_{(l)}(\theta),
\nfr{SMAT}
where $\theta$ is the rapidity difference of the incoming particles.
We now define the terms appearing in \SMAT.

The scalar factor $X^{ab}(\theta)$ is the minimal $S$-matrix associated
to $a_{m-1}$ [\Ref{TOD}]. It is usually expressed as
$$
X^{ab}(\theta)=\prod_{j=|a-b|+1\atop{\rm step}\ 2}^{b+a-1}{\sinh\left(
{\theta\over2}+i{\pi(j+1)\over2m}\right)\sinh\left({\theta\over2}+i{\pi
(j-1)\over2m}\right)\over\sinh\left({\theta\over2}-i{\pi(j+1)\over2m}\right)
\sinh\left({\theta\over2}-i{\pi(j-1)\over2m}\right)},
\efr
however for our purposes it is more useful to use the following
integral expression which is valid for physical rapidities
(i.e. ${\rm Im}(\theta)=0$):
$$
X^{ab}(\theta)=\exp\left\{i\pi\delta_{ab}-\int_{-\infty}^\infty{dx\over x}
e^{im\theta x/\pi}
\left[\hat A^{(m)}_{ab}(x)-\delta_{ab}\right]\right\},
\nfr{DEFX}
where the kernel, symmetric in its indices, is given by
$$
\hat A^{(m)}_{ab}(x)={2\sinh(ax)\sinh[(m-b)x]\cosh x
\over\sinh(mx)\sinh x},\qquad b\geq a.
\nfr{DEFA}

The other factors in \SMAT\
specify the non-trivial exchange of colour between the multiplets. They are
proportional to the RSOS $R$-matrices for the quantum group associated to
$a_{m-1}$ with deformation parameters equal to roots of unity:
$-\exp(-i\pi/(m+k))$ and $-\exp(-i\pi/(m+l))$,
respectively. The $S$-matrices describe a set of kinks interpolating
between a set of vacua which lie in the product of two highest weights of
$a_{m-1}$ subject to $\mu\cdot\vartheta\leq k$ and $\mu'\cdot\vartheta\leq
l$, respectively, where $\vartheta$ is the highest root.
We shall not require the explicit form of the $S$-matrix elements but only
their spectral decompositions.

If we choose the convention that $\widetilde S^{ab}_{(k)}(\theta)$ maps
$V_a\otimes V_b$ to $V_a\otimes V_b$ (so with no permutation of the
out-going vector spaces) then the spectral decomposition of the $S$-matrix
element for $b\geq a$ is
$$
{\widetilde
S}^{ab}_{(k)}(\theta)=\sigma^{ab}_{(k)}(\t)\sum_{j=0}^{{\rm min}(m-b,a)}
\rho^{ab}_j(\t){\Bbb P}_j,
\efr
where ${\Bbb P}_j$ is the quantum group
projector onto the representation with highest weight $\omega_{a-j}+
\omega_{b+j}$ that appears in the tensor product
$V_a\otimes V_b$ with the understanding that the projector
is zero if the highest weight does not satisfy $\mu\cdot
\vartheta\leq k$. In the above, $\sigma^{ab}_{(k)}(\theta)$ and
$\rho^{ab}_j(\theta)$ are scalar factors, symmetric in $a$ and $b$,
and with the normalization $\rho^{ab}_0(\theta)=1$, which
can be extracted from [\Ref{THII}]. For physical values of rapidity
and $b\geq a$ one can express them as
$$\sigma^{ab}_{(k)}(\theta)=
\exp \int_{-\infty}^{\infty}{dx\over x}
e^{im\theta x/\pi}{\sinh(ax)\sinh[(m+k-1)x]\sinh[(m-b)x]
\over\sinh[(m+k)x]\sinh(mx)\sinh x},
\efr
and for $j\geq1$
$$
\rho^{ab}_j(\theta)=c_j^{ab}\prod_{p=1}^j{\sinh
\left[{m\over2(m+k)}\left(\t+i{\pi\over m}(2p+b-a)\right)\right]\over
\sinh\left[{m\over2(m+k)}\left(\t-i{\pi\over m}(2p+b-a)\right)\right]},
\efr
where $c_j^{ab}$ is a constant whose exact form we shall not require.

The diagonal
$S$-matrix elements have the following important property:
$$
S^{aa}(0)=-P,
\nfr{ZER}
where $P$ is the permutation operator of the incoming states. This means that
the particles are of ``fermion-type'' to use the terminology of [\Ref{TBA}].

\chapter{The thermodynamic Bethe Ansatz}

The goal of the TBA is to extract information about a
integrable field theory from its factorizable $S$-matrix [\Ref{TBA}].
In particular one can probe the ultra-violet behaviour of the theory,
which will be some massless theory and hence conformally
invariant. In practice, it turns out that the central charge of the conformal
field theory is the most accessible piece of information.
The technique of the TBA has been dealt with at length elsewhere
(see [\Ref{TBA}] and references therein) and so
our {\it modus operandi\/} will be to only dwell on the details that are
peculiar to the problem of dealing with non-diagonal $S$-matrices.

Since particles are neither created nor destroyed it makes sense to consider
the thermodynamics of the one-dimensional gas of $N$ particles on a circle
with circumference $L$. Suppose that the $j^{\rm th}$
particle has rapidity $\theta_j$ and transforms in the representation
$a_j$ and hence has mass $m_j=m_{a_j}$.

The asymptotic wavefunction of the system, which is valid if the particles
are far apart apart compared with the interaction length ($\sim m_0^{-1}$)
has the familiar Bethe-Ansatz-type form
$$
\Psi(x_1,\ldots,x_N)=\exp\left(i\sum_{j=1}^Nx_jm_j\sinh\theta_j\right)
\sum_{Q\in S_N}\Theta(x_Q)\zeta(Q),
\efr
where the sum is over the permutations $Q=\{Q_1,\ldots,Q_N\}$ of
$\{1,2,\ldots,N\}$ with
$$
\Theta(x_Q)=\cases{1\qquad&if$\ x_{Q_1}<x_{Q_2}<\cdots<x_{Q_N}$\cr
0\qquad&otherwise,\cr}
\efr
and $\zeta(Q)$ is an element of the product of two copies of the
tensor product $V_{a_1}\otimes V_{a_2}
\otimes\cdots\otimes V_{a_N}$. The coefficients $\zeta(Q)$ are related
via the $S$-matrix in such a way that
if $Q$ differs from $Q'$ only by a permutation
of the $i^{\rm th}$ and $j^{\rm th}$ particles then
$$
\zeta(Q')=S^{a_ia_j}_{(k,l)}(\theta_i-\theta_j)\zeta(Q),\qquad i<j.
\efr
These relations determine all the $\zeta(Q)$ in terms of a reference value.

We now impose periodic boundary
conditions on the asymptotic wavefunction:
$$
\Psi(\ldots,x_j,\ldots)=\Psi(\ldots,x_j+L,\ldots),\qquad{\rm for}\
j=1,2,\ldots,N.
\efr
This implies that the wavefunction must be a simultaneous eigenvector
of the {\it transfer matrix\/} for bringing the $j^{\rm th}$ particle through
the rest and back to its starting point:
$$
\exp\left({im_jL\sinh\theta_j}\right)T(\t_j|\t_{j+1},\ldots,\t_N,\t_1,\ldots,
\t_{j-1})\Psi=\Psi,\qquad j=1,\ldots,N.
\nfr{EIGE}
The transfer matrix $T(\t_j|\t_{j+1},\ldots,\t_N,\t_1\ldots,\t_{j+1})$ is
constructed out the $S$-matrix in the following way [\Ref{FI}].
Consider the product
of $S$-matrix elements involving an auxiliary particle with rapidity $\t$ and
colour index $a$:
$$
T^a(\t)=S^{aa_1}(\t-\t_1)S^{aa_2}(\t-\t_2)\cdots S^{aa_N}(\t-\t_N),
\efr
where the labels $\{a_j\}$ and $\{\t_j\}$ are understood.
The transfer matrices are now given by
$$
T(\t_j|\t_{j+1},\ldots,\t_N,\t_1,\ldots,\t_{j-1})=-{\rm tr}_0
T^{a=a_j}(\t=\t_j),
\nfr{TRANMAT}
where the trace is over the representation
$a$ of the auxiliary particle. In order to prove \TRANMAT\ one uses the
fact \ZER\ that the diagonal $S$-matrix element at zero rapidity difference is
minus the permutation. The advantage of writing the transfer matrix
in this way is that the standard techniques pioneered in the field of
integrable lattice models may be wheeled-in to analyze its spectrum.
The only difference in the present context is that there are local
inhomogeneities specified by the rapidities $\{\t_j\}$ and the particles
carry varying colour indices $\{a_j\}$.

The fact that the $S$-matrix satisfies the Yang-Baxter equation guarantees
that the ${\rm tr}_0T^a(\t)$ commute for different values of $\t$ and $a$;
hence the wavefunction can indeed
be a simultaneous eigenvector of the transfer matrices for different $j$.
The diagonalization of the ${\rm tr}_0T^{a_j}(\t_j)$ can be accomplished via
standard Bethe Ansatz techniques. Recall that ${\rm tr}_0T^{a_j}(\t_j)$ acts
on two copies of
$V_{a_1}\otimes\cdots\otimes V_{a_N}$ (reflecting the product
form of the $S$-matrix) and is furthermore commutant to the
action of the quantum group implying that the eigenvectors correspond to
certain irreducible subspaces of $V_{a_1}\otimes\cdots\otimes
V_{a_N}$. The eigenvalues are a sum of terms and depend upon a set of
parameters which satisfy the Bethe Ansatz equations [\Ref{BRI},\Ref{DV}].
Fortunately, we do not require the exact expression
for the eigenvalues (which are only available in limited cases) rather
only the dominant term in the thermodynamic limit ($N\rightarrow\infty$).
Even so, this problem has not been solved in general and we must rely
on a conjecture [\Ref{BRI},\Ref{DV}] which reproduces all the known results.

The conjecture implies that the dominant contribution to the
eigenvalues of ${\rm tr}_0T^{a_j}(\t_j)$ is expressed as
$$\eqalign{
&\epsilon_j
\prod_{i=1\atop\neq j}^N\left[X^{a_ja_i}(\t_j-\t_i)\sigma^{a_ja_i}_{(k)}
(\t_j-\t_i)\sigma^{a_ja_i}_{(l)}(\t_j-\t_i)\right]\cr
&\times\prod_{\alpha=1}^{M_{a_j}}{\sinh\left[{m\over2(m+k)}\left(\theta_j-
u_\alpha^{(a_j)}+i{\pi\over m}\right)\right]
\over\sinh\left[{m\over2(m+k)}\left(\theta_j-
u_\alpha^{(a_j)}-i{\pi\over m}\right)\right]}
\prod_{\alpha=1}^{M'_{a_j}}{\sinh\left[{m\over2(m+l)}\left(\theta_j-
v_\alpha^{(a_j)}+i{\pi\over m}\right)\right]
\over\sinh\left[{m\over2(m+l)}\left(\theta_j-
v_\alpha^{(a_j)}-i{\pi\over m}\right)\right]},\cr}
\nfr{EIG}
where $\epsilon_j$ is a constant whose explicit form we shall not require.
The eigenvalues, for given whole numbers $M_a$ and $M'_a$,
$a=1,2,\ldots,m-1$ (which are subject to constraints to be discussed
shortly), are then
determined by the parameters $u_\alpha^{(a)}$ and $v_\alpha^{(a)}$
which are separately solutions of a set Bethe Ansatz equations:
$$\eqalign{
\prod_{j=1}^N&{\sinh\left[{m\over2(m+k)}\left(u_\alpha^{(a)}-\theta_j
+i{\pi\over m}\omega_{a_j}\cdot\alpha_a\right)\right]\over
\sinh\left[{m\over2(m+k)}\left(u_\alpha^{(a)}-\theta_j
-i{\pi\over m}\omega_{a_j}\cdot\alpha_a\right)\right]}\cr
&\qquad\qquad=\Omega^{(a)}_\alpha
\prod_{b=1}^{m-1}\prod_{\beta=1}^{M_b}{\sinh\left[{m\over2(m+k)}
\left(u_\alpha^{(a)}-u_\beta^{(b)}+i{\pi\over m}
\alpha_a\cdot\alpha_b\right)\right]
\over
\sinh\left[{m\over2(m+k)}
\left(u_\alpha^{(a)}-u_\beta^{(b)}-i{\pi\over m}
\alpha_a\cdot\alpha_b\right)\right]},\cr}
\nfr{BAE}
where $\Omega^{(a)}_\alpha$ is a phase
appropriate to the RSOS Bethe Ansatz, whose explicit form we shall not
require. The $v_\alpha^{(a)}$
satisfy an analogous set of equations with $k$ replaced by $l$ and $M_a$ by
$M'_a$. In \BAE\ the
$\alpha_a$'s are the simple roots of $a_{m-1}$ dual to the fundamental
weights: $\alpha_a\cdot\omega_b=\delta_{ab}$.

The eigenvalue \EIG\ is associated to a representation of the product
$a_{m-1}\times a_{m-1}$ with highest weight
$$
(\mu,\mu')=\left(\sum_{i=1}^N\omega_{a_i}-\sum_{a=1}^{m-1}M_a\alpha_a,
\sum_{i=1}^N\omega_{a_i}-\sum_{a=1}^{m-1}M'_a\alpha_a\right).
\nfr{HW}
The fact that $\mu$ and $\mu'$ have to be highest weights and furthermore
$\mu\cdot\vartheta\leq k$ and $\mu'\cdot\vartheta\leq l$ imposes restrictions
on the possible choices for $M_a$ and $M'_a$.

In the thermodynamic limit ($N\rightarrow\infty$ and $L\rightarrow\infty$)
the number of particles becomes large and the separation between the states
in rapidity-space becomes small, so that it
is meaningful to discuss the density of particles in rapidity-space
(per unit length of real space) with colour index $a$, a quantity
we denote $\sigma^a(\theta)$. As usual we also introduce $\tilde\sigma^a(\t)$
the density of unoccupied one-particle states (or holes) states
so that the density of one-particle states with colour index $a$ is then
$\sigma^a(\t)+\tilde\sigma^a(\t)$.
In taking the thermodynamic limit of the Bethe Ansatz equations \BAE\ we
have to employ the standard conjecture concerning the nature of the solutions
which dominate in the thermodynamic limit [\Ref{BRI}]. This is the string
hypothesis that maintains that the dominating solutions
are `$p$-strings' having the form
$$
u^{(a)}=u^{(a)}_0+i{\pi\over m}(p+1-2j),\qquad j=1,2,\ldots,p,
\efr
where the centre of the string $u^{(a)}_0\in{\Bbb R}$ and $p=1,2,\ldots,k$.
We then introduce the densities of the centres of occupied $p$-strings
and their associated
holes $r^a_p(\theta)$ and $\tilde r^a_p(\theta)$, respectively.
A similar picture holds for the Bethe Ansatz equations for the $v^{(a)}$,
requiring the introduction of $p$-string densities
$s^a_p(\t)$ and $\tilde s^a_p(\t)$, where in this case $p=1,2,
\ldots,l$. The $r$ and $s$ densities are often called {\it magnon\/}
densities.

The next step is to take the logarithm of the eigenvalue equation \EIGE\ and
then differentiate with respect to $\t_j$. In the thermodynamic limit the
discrete sums are replaced by integrals over the densities that we
introduced above and the resulting integral equations just depends on
the colour index $a_j$ and not on $j$ itself. In taking the thermodynamic limit
it is important that the particles are of fermionic-type. With this in mind,
the eigenvalue equations \EIGE\ become the coupled integral equations
$$
{m_a\over2\pi}\cosh\t+\gamma^{ab}_{(k,l)}\ast\sigma^b(\theta)
-a^{(m+k)}_p\ast r^a_p(\t)-a^{(m+l)}_p\ast s^a_p(\t)=\sigma^a(\t)+
\tilde\sigma^a(\t),
\nfr{INTE}
for $a=1,2,\ldots,m-1$ (repeated indices are summed)
where $\ast$ denotes the convolution $f\ast g(\t)=\int_{-\infty}^\infty
d\t'f(\t-\t')g(\t')$, and the kernels are found to be after some algebra
$$\eqalign{
&\gamma^{ab}_{(k,l)}(\t)=
{1\over2\pi i}{d\over d\t}\log\left[X^{ab}(\t)\sigma^{ab}_
{(k)}(\t)\sigma^{ab}_{(l)}(\t)\right]\cr
=&\delta(\t)\delta_{ab}
-{m\over\pi}\int_{-\infty}^{\infty}{dx\over2\pi}
e^{im\t x/\pi}{\sinh(ax)\sinh[(m-b)x]\sinh[(2m+k+l)x]
\over\sinh(mx)\sinh[(m+k)x]\sinh[(m+l)x]}\cr
=&\delta(\t)\delta_{ab}-\left[A_{m+l,m+l}^{(2m+k+l)}\right]^{-1}\ast
A^{(m)}_{ab}(\t),\cr}
\nfr{BLI}
and
$$\eqalign{
a_p^{(m+k)}(\t)=&{1\over2\pi i}{d\over d\t}\sum_{j=1}^p\log
{\sinh\left[{m\over2(m+k)}\left(\t-i{\pi\over m}(p+2-2j)\right)\right]\over
\sinh\left[{m\over2(m+k)}\left(\t+i{\pi\over m}(p+2-2j)\right)\right]}\cr
=&{m\over\pi}\int_{-\infty}^\infty{dx\over2\pi}e^{im\t x/\pi}
{\sinh[(m+k-p)x]\over\sinh[(m+k)x]},\cr}
\efr
where $A^{(m)}_{ab}(\t)$ is defined in terms of $\hat A^{(m)}_{ab}(x)$ in
\DEFA\ using the general definition of the Fourier transform
$$
f(\t)={m\over\pi}\int_{-\infty}^\infty{dx\over2\pi}e^{im\t x/\pi}
\hat f(x),
\efr
and we define
$$
\left[f\right]^{-1}(\t)={m\over\pi}\int_{-\infty}^\infty{dx\over2\pi}
e^{im\t x/\pi}{1\over\hat f(x)}.
\efr

In a similar way, the Bethe Ansatz equations yield in the
thermodynamic limit [\Ref{BRI}]
$$\eqalign{
\tilde r^a_p(\t)+A^{(m+k)}_{pq}\ast K^{(m)}_{ab}\ast r^b_q(\t)=&
a^{(m+k)}_p\ast\sigma^a(\t),\quad p=1,2,\ldots,k\cr
\tilde s^a_p(\t)+A^{(m+l)}_{pq}\ast K^{(m)}_{ab}\ast s^b_q(\t)=&
a^{(m+l)}_p\ast\sigma^a(\t),\quad p=1,2,\ldots,l,\cr}
\nfr{BAEE}
where
$$
K^{(m)}_{ab}(\t)={m\over\pi}\int_{-\infty}^\infty{dx\over2\pi}e^{im\t x/\pi}
\left({\delta_{ab}-{1\over2\cosh x}I^{(m)}_{ab}}\right),
\efr
and $I^{(m)}$ is the incidence matrix of the $a_{m-1}$ Dynkin
diagram. It can be shown [\Ref{BRII}] that $K^{(m)}_{ab}(\t)$
is actually the inverse of
$A^{(m)}_{ab}(\t)$ in the sense that $K^{(m)}_{ab}\ast A^{(m)}_{bc}(\t)=
\delta(\t)\delta_{ac}$.

Using \BLI\ equation \INTE\ can be rewritten as
$$
\tilde\sigma^a(\t)+\left[A^{(2m+k+l)}_{m+l,m+l}\right]^{-1}\ast
A^{(m)}_{ab}\ast\sigma^b(\t)+a^{(m+k)}_p\ast r^a_p(\t)
+a^{(m+l)}_p\ast s^a_p(\t)={m_a\over2\pi}\cosh\t.
\nfr{INTT}

The equations can be simplified because we shall argue that the
$k$-strings in the $r$-system and the $l$-strings in the $s$-system do not
contribute in the thermodynamic limit. Consider the
zero-mode of the $r$-system \BAEE\ at $p=k$:
$$
\hat{\tilde r}^a_k(0)+\sum_{b=1}^{m-1}
\sum_{q=1}^k{qm\over m+k}C_{ab}\hat r_q^b(0)=
{m\over m+k}\hat\sigma^a(0),
\nfr{ZM}
where $C_{ab}$ is the Cartan matrix of $a_{m-1}$.
In the thermodynamic limit the highest weight \HW\ of the eigenvector
becomes
$$
\mu=L\sum_{a=1}^{m-1}\left(\hat\sigma^a(0)\omega_a-\sum_{q=1}^kq\hat r^a_q(0)
\alpha_a\right)
=L{m+k\over m}\sum_{a=1}^{m-1}\hat{\tilde r}^a_k(0)\omega_a,
\efr
by \ZM. But we know that $\mu$ is bounded by $\mu\cdot\vartheta\leq k$
and so the number of $k$-string holes of $u^{(a)}$ goes to zero in the
thermodynamic limit and so we can safely take $\tilde r^a_k(\t)=0$ and
similarly $\tilde s^a_l(\t)=0$ in \BAEE\ and \INTT, for $a=1,2,\ldots,m-1$.
With $\tilde r^a_k(\t)=0$ one can eliminate $r^a_k(\t)$ via
$$
r^a_k(\t)=-\left[A^{(m+k)}_{kk}\right]^{-1}\ast A^{(m+k)}_{kp}\ast r^a_p(\t)
+\left[A^{(m+k)}_{kk}\right]^{-1}\ast a^{(m+k)}_k\ast A^{(m)}_{ab}\ast
\sigma^b(\t),
\efr
and similarly for $s^a_l(\t)$. Substituting for $r^a_k(\t)$ and $s^a_l(\t)$ in
\INTT\ and after some algebra (in particular the identities in appendix A of
[\Ref{BRII}] are helpful) leads finally to
$$
\tilde\sigma^a(\t)+\left[A^{(k+l)}_{ll}\right]^{-1}\ast A^{(m)}_{ab}\ast
\sigma^b(\t)+a^{(k)}_p\ast r^a_p(\t)+a^{(l)}_p\ast s^a_p(\t)=
{m_a\over2\pi}\cosh\t.
\nfr{INTF}
In a similar way \BAEE\ becomes
$$\eqalign{
\tilde r^a_p(\t)+A^{(k)}_{pq}\ast K_{ab}^{(m)}\ast r^b_q(\t)&=a^{(k)}_p\ast
\sigma^a(\t)\cr
\tilde s^a_p(\t)+A^{(l)}_{pq}\ast K_{ab}^{(m)}\ast r^b_q(\t)&=a^{(l)}_p\ast
\sigma^a(\t),\cr}
\nfr{BAEF}
where the string labels on $r$ now run from $1$ to $k-1$ and those on $s$
run from $1$ to $l-1$. Notice that \INTF\ and \BAEF\ are identical to the
Bethe Ansatz equations for the integrable lattice model in regime I of
[\Ref{BRI},\Ref{BRII}].

The equations can be amalgamated by defining
$$
\rho^a_p(\t)=\cases{\tilde s^a_{l-p}(\t)\qquad&$p=1,2,\ldots,l-1$\cr
\sigma^a(\t)&$p=l$\cr \tilde r^a_{p-l}(\t)&$p=l+1,l+2,\ldots,l+k-1$,\cr}
\efr
and
$$
\tilde\rho^a_p(\t)=\cases{s^a_{l-p}(\t)\qquad&$p=1,2,\ldots,l-1$\cr
\tilde\sigma^a(\t)&$p=l$\cr r^a_{p-l}(\t)&$p=l+1,l+2,\ldots,l+k-1.$\cr}
\efr
After some algebra, and using the fact that $K^{(k+l)}_{pq}(\t)$ is the
inverse of $A^{(k+l)}_{pq}(\t)$, it follows that \INTF\ and \BAEF\ are
subsumed by
$$
\tilde\rho^a_p(\t)+K^{(k+l)}_{pq}\ast A_{ab}^{(m)}\ast\rho^b_q(\t)=
\delta_{pl}{m_a\over2\pi}\cosh\t,
\nfr{AMB}
where the string label on $\rho$ runs from $1$ to $k+l-1$.

We are now in a position to discuss the thermodynamics of the system at a
temperature $T$. To this end we note that the energy per unit length
of a solution to the equations is
$$
{\cal E}=\int_{-\infty}^\infty d\t\ \sigma^a(\t)m_a\cosh\t\equiv
\int_{-\infty}^\infty d\t\ \rho^a_l(\t)m_a\cosh\t,
\efr
where as usual there is an implied sum over the repeated index. The entropy
per unit length of a solution is given by
$$
{\cal S}=\int_{-\infty}^\infty d\t\left\{\left(\rho^a_p+\tilde\rho^a_p\right)
\log
\left(\rho^a_p+\tilde\rho^a_p\right)-\rho^a_p\log\rho^a_p-\tilde\rho^a_p
\log\tilde\rho^a_p\right\},
\efr
which is the form appropriate to a set of particles of fermionic-type.
The next stage is to minimize the free energy per unit length
${\cal F}={\cal E}-T{\cal S}$ subject to the constraint \AMB. The resulting
equation for the densities is
$$
{1\over T}\delta_{pl}m_a\cosh\t=\epsilon^a_p(\t)+L^a_p(\t)
-K^{(k+l)}_{pq}\ast A^{(m)}_{ab}\ast L^b_q(\t),
\nfr{THE}
where we used the fact that $K^{(k+l)}_{pq}(\t)$ is the inverse of $A^{(k+l)}
_{pq}(\t)$ and we have defined
$$
{\tilde\rho^a_p(\t)\over\rho^a_p(\t)}=\exp\epsilon^a_p(\t),
\qquad L^a_p(\t)=\log\left(1+e^{-\epsilon^a_p(\t)}\right).
\efr
Once these equations are solved the free energy per unit length is given by
$$
{\cal F}=-{T\over2\pi}\int_{-\infty}^\infty d\t\ L^a_l(\t)m_a\cosh\t.
\efr
This completes the derivation of the TBA system resulting from the
trigonometric $S$-matrices.

To make contact with [\Ref{RAV}] one defines
$$
{1\over2\pi}\phi^{ab}(\t)={1\over2\pi i}{d\over d\t}\log X^{ab}(\t)=
\delta_{ab}\delta(\t)-A^{(m)}_{ab}(\t),
\efr
and $\psi^{ab}(\t)$ via its Fourier transform:
$$
{1\over2\pi}\hat\psi^{ab}(x)={1\over2\cosh x}\left({\delta_{ab}-{1\over2\pi}
\hat\phi^{ab}(x)}\right).
\efr
In terms of these quantities \THE\ becomes precisely equation (2) of
the first reference of [\Ref{RAV}] (except for the error in sign):
$$
{1\over T}\delta_{pl}m_a\cosh\t=\epsilon^a_p(\t)
+{1\over2\pi}\left({\phi^{ab}\ast L^b_p(\t)+I^{(k+l)}_{pq}\psi^{ab}\ast
L^b_q(\t)}\right),
\nfr{RTBA}
where $I^{(k+l)}$ is the incidence matrix of the $a_{k+l-1}$ Dynkin diagram.

In the limit $k,l\rightarrow\infty$ the TBA system is also identical to that
proposed for the $SU(m)$ principle chiral model [\Ref{PCM}]
reflecting the fact that the trigonometric $S$-matrix also reduces to
the conjectured $S$-matrix of this
model in that limit. Also when $k=l=1$ \RTBA\ reduces to that TBA system for
the $a_{m-1}$ purely elastic scattering theory as it should since in
this case the $S$-matrix collapses to the scalar factor
$S^{ab}_{(1,1)}(\t)=X^{ab}(\t)$.

\chapter{Discussion}

We have shown how the TBA systems for the algebras $a_{m-1}$ are
associated to the trigonometric $S$-matrices. One point which is worth
emphasizing is that in going from the $S$-matrices to the TBA equations
one uses the expression for dominant contribution to the eigenvalues
of the transfer matrices in thermodynamic limit. It seems, therefore, that it
is not possible to reverse the process and deduce the $S$-matrix elements
from the TBA equations. Unfortunately, this seems to preclude a
deduction of the $S$-matrices
associated to other algebras (only the full $S$-matrices of the
$a_{m-1}$ and $c_m$ theories are known [\Ref{THII}]) from the TBA equations.

Using the analysis of [\Ref{RAV}] we can now investigate the UV-limit of
the scattering theory defined in section 2. The UV limit corresponds
taking to
$T\rightarrow\infty$. The free energy can be interpreted as $R^{-1}$ times the
Casimir energy of the quantum field theory on a periodic strip with period
$R=1/T$. It is known from the theory of finite-size effects
that the Casimir energy of the periodic strip in the UV-linit
gives the central charge $c$ of the resulting conformal field theory via
$$
\lim_{R\rightarrow0}R^2{\cal F}(R)=-{\pi c\over6}.
\efr
The central charge that one finds is that of the coset $g_k\times
g_l/g_{k+l}$ where $g=a_{m-1}$.

The construction of the TBA system has been based on the $S$-matrix of (2.2).
It is possible to write down an $S$-matrix
having the more general form
$$
S^{ab}_{(k_1,k_2,\ldots,k_s)}(\theta)=X^{ab}(\theta){\widetilde S}^{ab}_{(k_1)}
(\t)\otimes{\widetilde S}^{ab}_{(k_2)}(\t)
\otimes\cdots\otimes{\widetilde S}^{ab}_{(k_s)}(\theta),
\efr
for a set of natural numbers $\{k_1,k_2,\ldots,k_s\}$,
where the particles now transform in the representations $V_a^{\otimes\,s}$.
It would be interesting to calculate the central charge of the UV-limit
of this more general theory.

\references

\beginref
\Rref{RSOS}{M. Jimbo, T. Miwa and M. Okado, Comm. Math. Phys. {\bf116}
(1988) 507}
\Rref{TBA}{A.B. Zamolodchikov, Nucl. Phys. {\bf B342} (1990) 695\newline
T.R. Klassen and E. Melzer, Nucl. Phys. {\bf B338} (1990) 485; Nucl. Phys.
{\bf B350} (1991) 635\newline
M.J. Martins, Phys. Lett. {\bf B240} (1990) 404}
\Rref{ND}{A.B. Zamolodchikov, Nucl. Phys. {\bf B385} (1991) 497; Nucl. Phys.
{\bf B385} (1991) 524; Nucl. Phys. {\bf B366} (1991) 122\newline
P. Fendley and H. Saleur, {\bf B388} (1992) 609}
\Rref{RAV}{F. Ravanini, Phys. Lett. {\bf B282} (1992) 73\newline
M.J. Martins, Phys. Lett. {\bf B277} (1992) 301}
\Rref{BRI}{V.V. Bazhanov and N. Reshetikhin, J. Phys. {\bf A23} (1990) 1477}
\Rref{BRII}{V.V. Bazhanov and N. Reshetikhin, Prog. Theor. Phys. Suppl.
{\bf102} (1990) 301}
\Rref{VF}{H.J. de Vega and V.A. Fateev, Int. J. Mod. Phys. {\bf A6} (1991)
3221}
\Rref{THI}{T.J. Hollowood, Int. J. Mod. Phys. {\bf A8} (1993) 947}
\Rref{THII}{T.J. Hollowood, {\sl The analytic structure of trigonometric
S-matrices\/}, CERN preprint CERN-TH.6888/93, {\tt hep-th/9305042}}
\Rref{PCM}{E. Ogievetsky, P. Wiegmann and N. Reshetikhin, Nucl. Phys. {\bf
B280} (1987) 45\newline P.Wiegmann, Phys. Lett.{\bf B141} (1984) 217}
\Rref{TOD}{A.E. Arinstein, V.A. Fateev and A.B. Zamolodchikov, Phys. Lett.
{\bf B87} (1979) 389}
\Rref{FI}{P. Fendley and K. Intriligator, Nucl. Phys. {\bf B372} (1992) 533}
\Rref{DV}{H.J. de Vega, Int. J. Mod. Phys. {\bf A4} (1989) 2371; Int. J. Mod.
Phys. {\bf A5} (1990) 1611 \newline
N. Reshetikhin and P.B. Wiegmann, Phys. Lett. {\bf B189} (1987) 125}
\Rref{ABL}{A. Ahn, D. Bernard and A. LeClair, Nucl. Phys. {\bf B346} (1990)
409}
\endref
\ciao